\documentclass[prl,twocolumn,a4paper,floatfix]{revtex4}
\usepackage{graphicx}
\usepackage{bbm}
\usepackage{amsmath}

\begin{document}

\title{Collective non-Abelian instabilities in a melting Color Glass Condensate}

\author{Paul Romatschke$^{\rm a}$
        and
        Raju Venugopalan$^{\rm b}$
        }

\affiliation{a\ Fakult\"{a}t f\"{u}r Physik, Universit\"{a}t Bielefeld, 33615 Bielefeld, Germany\\ 
 b\ Physics Department, Brookhaven National Laboratory, Upton,
 N.Y. 11973, U.S.A.\\
}

\date{\today}

\begin{abstract}
We present first results for 3+1-D simulations of SU(2) Yang-Mills equations for matter expanding into the vacuum after a heavy ion collision. Violations of 
boost invariance cause a non-Abelian Weibel instability leading soft modes to grow with proper time $\tau$ as $\exp(\Gamma \sqrt{g^2\mu \tau})$, where $g^2\mu$ is a scale arising from the saturation of gluons in the nuclear wavefunction. The scale for the growth rate $\Gamma$ is set by a 
plasmon mass, defined as $\omega_{\rm pl}= \kappa_0\,\sqrt{\frac{g^2\mu}{\tau}}$, generated
dynamically in the collision. We compare the numerical ratio $\Gamma/\kappa_0$  to the corresponding value predicted by the Hard Thermal Loop formalism for anisotropic plasmas.

\end{abstract}

%\pacs{25.75.-q, 24.10.-i, 24.85.+p}

\maketitle

Experiments at the Relativistic Heavy Ion Collider (RHIC) at Brookhaven National Laboratory indicate  
that a thermalized Quark Gluon Plasma (QGP) may be formed in ultrarelativistic collisions of beams of gold ions~\cite{whitepaper}.  Phenomenological analyses of RHIC data point to thermalization of matter on early time scales of order 1 fm/c after the collision~\cite{hydro}. Understanding how wavefunctions of high energy nuclei decohere completely on short time scales to form a thermalized QGP 
is an outstanding theoretical puzzle of great interest. 

At central rapidities, where thermalization is most likely,  the wavefunctions of colliding nuclei are 
dominated by small x modes which, due to their large occupation numbers, are described 
as classical fields~\cite{MV}.  These fields, and their evolution with energy, 
can be computed in a classical effective field theory called the Color Glass Condensate (CGC)~\cite{CGC} characterized by  a semi--hard scale, $Q_s(x) >> \Lambda_{\rm QCD}$. This ``saturation" scale ensures that the dynamics can in principle be understood in weak coupling; it is 
estimated~\cite{CGC} to be $Q_s\approx 1.4$ GeV for RHIC energies and $Q_s\approx 2.2$ GeV at Large Hadron Collider (LHC) energies of $\sqrt{s}=5.5$ TeV/nucleon.   The initial conditions for the collision can be expressed self-consistently in terms of the computed classical fields of each of the nuclei~\cite{KMW}.

The classical dynamics of  fields produced in the collision is approximately boost invariant. Assuming it to be strictly so, the resulting 2+1 dimensional Yang-Mills equations generating the space-time dynamics of fields (the ``melting of the CGC") was investigated numerically, and the energy and number distributions of the classically produced gluons computed~\cite{gluodynamics}. On proper time scales 
$\tau \sim 1/Q_s$, the energy density behaves as $\varepsilon\sim 1/\tau$, suggesting free streaming of gluons in the transverse plane at these 
early times. In their ``bottom up" scenario, Baier et al. suggested that thermalization is a consequence of subsequent re-scattering of on-shell gluons by $2\leftrightarrow 2$~\cite{MuellerBjRV} and $2\leftrightarrow 3$ processes~\cite{Wong,BMSS}. The bottom 
up estimates give $\tau_{\rm therm.} \sim \frac{1}{\alpha_S^{13/5} Q_s}$. This thermalization time scale, at RHIC energies, is 
significantly larger than 1 fm/c. 

Scattering rates in the bottom up scenario are controlled by the Debye mass, which sets the range of particle interactions. Interestingly, the Debye 
mass squared can change sign for anisotropic ``CGC like" initial
conditions, giving rise to an imaginary component in the gluon
dispersion relation~\cite{MikePaul}. This corresponds to the non-Abelian analog of the collective
Weibel instability~\cite{Weibel, Stanislav,ALM} leading to an  
exponential growth of soft modes. This phenomenon may 
invalidate or at least modify~\cite{MSW} bottom up thermalization. Recently, several numerical studies have explored the behavior of instabilities in the finite temperature Hard Thermal Loop (HTL) formalism~\cite{BraatenPisarski} extended to anisotropic non-Abelian plasmas~\cite{HTL_numerics}. These 
studies find exponentially growing soft modes. However, the most recent detailed 3+1-D simulations find that the exponential modes saturate 
at late times when non-linear self-interactions become important. Also interesting is a subsequent phenomenon, analogous to Kolmogorov cascading in turbulent plasmas, where energy is transferred back to hard modes~\cite{ArnoldMoore}. While all these studies consider 
anisotropic particle distributions {\it a la} the CGC, neither the expansion of the system nor the specific CGC dynamics following from physics of the 
small x modes is implemented. 

We present in this letter, for an SU(2) Yang-Mills theory, a first 3+1-dimensional numerical study of non-Abelian collective instabilities generated in the expanding CGC. The three 
spatial dimensions here are the two transverse co-ordinates $x_\perp$, the space time rapidity $\eta = \frac{1}{2}\ln(\frac{t+z}{t-z})$. 
The fourth dimension is given by  the proper time  $\tau=\sqrt{t^2 -z^2}$. The strict boost invariance of fields (their lack of dependence on $\eta$) in Ref.~\cite{gluodynamics} is relaxed~\footnote{It is essential to distinguish ``strict" boost invariance of fields from that  
of single particle distributions. The latter can be preserved even if the former is violated.}. Violations of boost invariance arise from finite 
energy constraints and from small x quantum corrections. The magnitude of the violations from the former are small. The 
latter are significantly larger and are of order unity over rapidity
intervals of order $Y\sim 1/\alpha_S$. We will only consider small
violations of boost invariance here;  as a consequence, 
the gluon field configurations are highly anisotropic in momentum space. 
Kinetic theory studies ~\cite{Stanislav,MikePaul,ALM} show that anisotropic distributions of hard modes rather than details of
their dynamics drive the non-Abelian Weibel instability. 
Our initial conditions are therefore similar, in this respect, because
the  high momentum  anisotropically distributed hard modes of the field with 
$k_\perp \sim Q_s$ play the role of ``hard'' particles in their
coupling to the soft modes with $k_\perp \approx \tau^{-1} k_\eta <<
Q_s$. 
The effects of large amplitude violations of boost invariance will be commented on briefly and discussed further in a follow up to this work~\cite{PaulRaju2}.

In $A^\tau=0$ gauge, initial conditions that violate strict boost invariance for the dynamical fields $A_\perp, A_\eta$ and their canonically conjugate momenta $E_\perp$, $E_\eta$ are 
\begin{equation}
A_i = {\cal A}_i\,\,;\,\, A_\eta = 0\,\,; \,\, E_i = \delta E_i\,\,; \,\, E_\eta = {\cal E}_\eta + \delta E_\eta \, ,\nonumber 
\label{eq:b_ic}
\end{equation}
where, for each configuration of the color charges of the two nuclei, ${\cal A}_i\equiv {\cal A}_i(x_\perp)$, $A_\eta =0$, $E_i =0$, ${\cal E}_\eta \equiv {\cal E}_\eta(x_\perp)$ are the boost invariant initial conditions studied in previous simulations~\cite{gluodynamics}.  The 
rapidity dependent functions $\delta E_i$ and $\delta E_\eta$ are constructed to satisfy Gauss's law, $D_i \delta E_i + D_\eta E_\eta =0$, at the 
initial time $\tau=\tau_0$. For each $\delta E_i$, random configurations $\delta {\bar 
E}_i (x_\perp)$ are drawn, satisfying $\langle \delta {\bar E}_i (x_\perp)\,\delta {\bar E}_i (y_\perp)\rangle = \delta^{(2)}(x_\perp - y_\perp)$. These 
random configurations are multiplied by another random function $f(\eta) = \partial_\eta F(\eta)$. One obtains, at $\tau=\tau_0$, 
%\begin{equation}
$$\delta E_i(x_\perp,\eta) = f(\eta)\,\delta{\bar E}_i(x_\perp)\,\,; \,\, \delta E_\eta = - F(\eta)\, D_i\,\delta {\bar E}_i(x_\perp)\, .
%\label{eq:2}
%\end{equation}
$$
The fields $F(\eta)$ are Gaussian white noise distributed, $\langle F(\eta)\,F(\eta^\prime)\rangle = \Delta^2\,\delta(\eta-\eta^\prime)$, with the 
amplitude of violations of boost invariance governed by the parameter $\Delta$. These initial conditions are not unique. Their virtue is 
that Gauss' law is manifest and periodic boundary conditions can be applied in the $\eta$ direction. A broader class of 
initial conditions will be considered in future.

Our numerical procedure is as follows~\footnote{This procedure is
similar to that of Ref.~\cite{gluodynamics}. Further details will be 
presented in Ref.~\cite{PaulRaju2}.}. The small x classical fields before the collision are determined from their respective classical color source 
densities by solving Poisson's equations~\cite{gluodynamics}. These color charge distributions are Gaussian distributed with a variance $g^2\mu$. This  momentum scale (closely related and similar in magnitude~\footnote{In the classical theory of the CGC, $Q_s^2 = g^4 \mu^2 N_c/2\pi \cdot \ln(g^2\mu/\Lambda)$, where 
$\Lambda$ is an infrared scale of order $\Lambda_{\rm QCD}$.}  to $Q_s$)  and the size of the system $L$  are the physical dimensionful scales 
in the problem.  For periodic boundary conditions, $L^2 = \pi R^2$, where $R$ is the radius of the nuclei. For each configuration of color charges, the initial conditions are determined, and an adaptive leap-frog algorithm evolves Hamilton's equations for the dynamical fields and their canonically conjugate momenta on a discretized space-time lattice. Physical results, obtained by averaging the results over 
all color configurations of the sources, are expressed in terms of $g^2\mu$ and the dimensionless combination $g^2 \mu L$. Higher energies and/or larger nuclei correspond to larger values of $g^2\mu L$.

The lattice parameters, in dimensionless units, are a) $N_\perp$ and $N_\eta$, the number of lattice sites in the $x_\perp$ and $\eta$ directions respectively; b) $g^2\mu\,a_\perp$ and $a_\eta$, the respective lattice spacings;  c) $\tau_0/a_\perp$ and $\delta\tau$, the time at which the simulations are initiated and the stepping size 
respectively; and finally d) $\Delta$, the initial size of the rapidity fluctuations. The continuum limit is obtained by holding the physical combinations 
$g^2\mu\,a_\perp\,N_\perp = g^2 \mu L$ and $a_\eta\,N_\eta = L_\eta$ fixed while sending $\delta \tau$, $g^2\mu\,a_\perp$ and $a_\eta$ to zero. 
For this study, we pick $L_\eta = 1.6$ units of rapidity. Variations of $L_\eta$ will be commented on later.
The magnitude of violations of boost invariance, as represented by $\Delta$, is a physical quantity;  here, we study results for small values of $\Delta\sim 10^{-11}-10^{-8}$.
The initial time is chosen to ensure that for $\Delta=0$, we recover earlier 2+1-D results; we set $\tau_0 =0.05\,a_\perp$. Our 
results are insensitive to variations that are a factor of 2 larger or smaller than this choice. 

\begin{figure}
\includegraphics[width=0.6\linewidth,angle=-90]{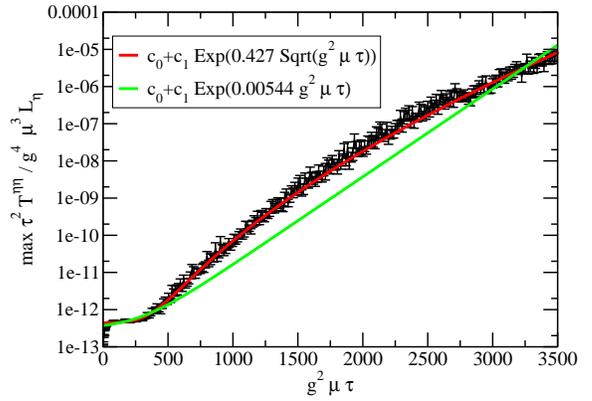}
\caption{The maximum Fourier mode amplitudes of $\tau^2 T^{\eta\eta}$ 
for $g^2\mu L=67.9$, $N_\perp=N_\eta=64$, $N_\eta a_\eta=1.6$.
Also shown are best fits with $\exp{\tau}$ and $\exp{\sqrt{\tau}}$
behavior. The former is clearly ruled out by the data.}
\label{fig:maxFM}
\end{figure}

To study the growth rate of instabilities due to violations of boost invariance, we define $$
{\tilde T}^{\mu\nu}(k_\eta,k_\perp=0) = \int d\eta \,\exp(i\eta\,k_\eta) \langle T^{\mu\nu}(x_\perp,\eta)\rangle_{\perp,\rho} \, ,$$
%\label{eq:ftTeta}
%\end{equation}
where $T^{\mu\nu}$ denote components of the stress energy tensor and $\langle \rangle_{\perp,\rho}$ denotes an average over the transverse co-ordinates and over all color charge configurations respectively. We will look in particular at this Fourier transform for the longitudinal pressure $\tau^2\, T^{\eta\eta}$. The magnitude of this quantity is a useful measure of isotropization; momentum distributions are isotropic when $2\,\tau^2 T^{\eta\eta}/
(T^{xx}+T^{yy})\rightarrow 1$. In Fig.~\ref{fig:maxFM}, we plot the maximal value of $\tau^2 {\tilde T}^{\eta\eta}$ at each time step, as a function of 
$g^2\mu \tau$. The data are for a $64^3$ lattice and correspond to
$g^2\mu L = 67.9$ and $L_\eta = 1.6$. The maximal value remains nearly
constant until $g^2\mu \tau \approx 250$, beyond which it grows
rapidly. A best fit to the functional form $c_0 + c_1 \exp({c_2
\tau^{c_3}})$ gives $c_2 =0.427\pm 0.01$ for $c_3 = 0.5$; the
coefficients 
$c_0$, $c_1$ are small numbers proportional to the initial seed. It is clear from Fig.~\ref{fig:maxFM} that the form $\exp(\Gamma \sqrt{g^2\mu \tau})$ is 
preferred to a fit with an exponential growth in $\tau$. 

We digress briefly to note that the relevant time scales in
Fig. \ref{fig:maxFM}, for RHIC
collisions where $g^2\mu \approx 1$ GeV, are too large to be
interest. These however correspond to the time necessary for 
extremely small violations of boost invariance to become
visible. The precise behavior of non-Abelian Weibel instabilities in a
melting CGC are more transparent for small initial seeds. 
\begin{figure}
\includegraphics[width=0.6\linewidth,angle=-90]{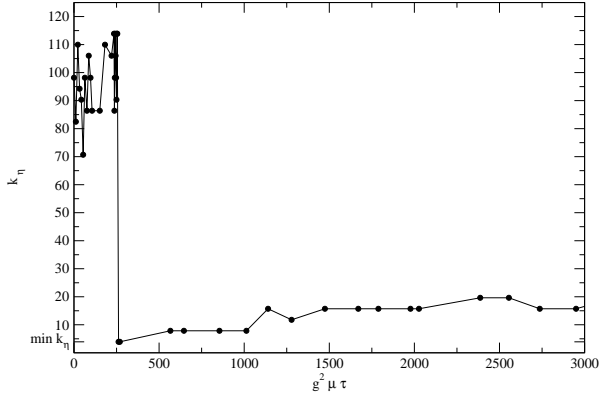}
\caption{The mode number $k_\eta$ for which the maximum amplitude
of $\tau^2 T^{\eta\eta}$ occurs (see Fig.\ref{fig:maxFM});
again for $g^2\mu L=67.9$, $N_\perp=N_\eta=64$, $N_\eta a_\eta=1.6$.
$k_{\rm min}$ denotes the smallest $k_\eta$ mode that could be excited
in the lattice simulation.}
\label{fig:max_keta}
\end{figure}

In Fig.~\ref{fig:max_keta}, we plot the most unstable $k_\eta$ mode,
corresponding to the maximal $\tau^2 {\tilde T}^{\eta\eta}$ in
Fig.~\ref{fig:maxFM}, as a function of $g^2\mu \tau$. Our result
suggests that the soft $k_\eta$ modes are sensitive to the non-Abelian Weibel
instability while the hard modes are not. Further, the maximal mode number $k_\eta$ 
grows as a function of time. 
The amplitude of the soft modes dominates the spectrum for 
$g^2\mu\tau> 250$. This timescale, however, is weakly dependent on our
choice of $L_\eta=1.6$;  raising 
$L_\eta$ to $L_\eta=3.2$ lowers the timescale for the onset of growth
by some $20\%$, while the extracted growth rate stays the same.
The large volume limit will be examined further in an
upcoming work \cite{PaulRaju2}.

We now turn to the correspondence between our results and the HTL formalism for anisotropic plasmas. In the latter, the maximal Fourier amplitudes of components of the stress-energy tensor 
grow as $\exp(2\gamma\tau)$, where the growth rate $\gamma$ satisfies the 
relation $\gamma=\sqrt{\frac{1}{2}}\,m_\infty$ for maximally anisotropic particle distributions~\cite{ALM}. Here,  
$$ m_\infty^2 =   g^2 N_c\,\int {d^3 p \over (2\pi)^3} {f({\vec p})\over p}\,,$$ where $f$ is the anisotropic single particle distribution of the hard modes. For both isotropic as well as anistropic distributions
(within the model used in \cite{MikePaul},
~\footnote{\label{HTLnote} Averaging over longitudinal and transverse
modes (see the 2nd paper by Rebhan et al. in
Ref.~\cite{HTL_numerics}), one obtains 
$\omega_{\rm pl}^2 = \frac{m^2}{3 \sqrt{\xi}}{\rm arctan}\sqrt{\xi}$ and $m_\infty^2=\frac{m^2}{2 \sqrt{\xi}} {\rm arctan} \sqrt{\xi}$, 
where $\xi$ denotes the strength of the anisotropy and $m^2$ is a soft scale
proportional to the Debye mass.  Therefore, $\omega_{\rm pl}^2 = \frac{2}{3} m_\infty^2$.})
$m_\infty^2 = \frac{3}{2}\,\omega_{\rm pl}^2$. 

\begin{figure}
\includegraphics[width=0.6\linewidth,angle=-90]{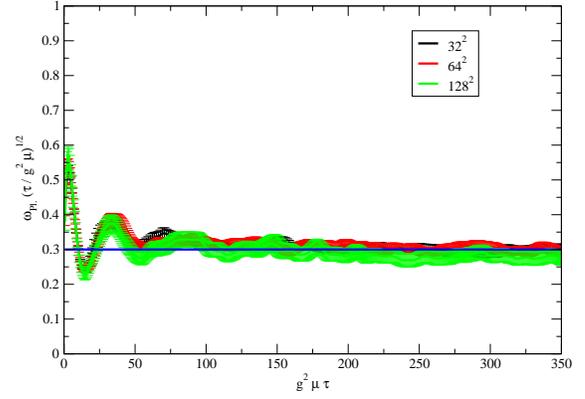}
\caption{Time evolution of $\omega_{\rm pl.}$, for
fixed $g^2 \mu L=22.6$ and lattice spacings
$g^2 \mu a_\perp=0.707,0.354,0.177$ ($N_\perp=32,64,128$), respectively.}
\label{fig:plasmon_tau}
\end{figure}

\begin{figure}
\vspace*{0.5cm}
\includegraphics[width=0.6\linewidth,angle=-90]{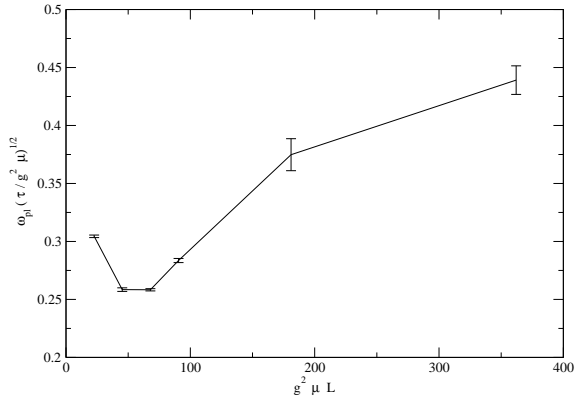}
\caption{Dependence of  $\omega_{\rm pl.}$ with respect to $g^2 \mu
L$; $g^2 \mu a_\perp=0.707$ for all $g^2 \mu L$ except $g^2 \mu L=67.9$, for
which it is $g^2 \mu a_\perp=1.06$.}
\label{fig:plasmon_g2muL}
\end{figure}

Interestingly, in close analogy to the HTL case, the $\exp(\Gamma \sqrt{g^2\mu \tau})$ growth of
the unstable soft modes in our case is closely related to a mass gap generated by the highly non-linear dynamics of soft modes in the expanding system. As in Ref.~\cite{gluodynamics}, fixing the residual gauge freedom 
$\nabla_\perp \cdot A_\perp$, the gluon dispersion is given by 
relation \footnote{The effect of small longitudinal fluctuations
on transverse quantities should be rather small. We therefore extract $\omega({\bf k_\perp})$, for computational convenience, 
from a 2+1-D simulation.}
$$ \omega({\bf k}_\perp) = {1\over \tau}\sqrt{{{\rm Tr} \left[ E_i({\bf k}_\perp E_i(-{\bf k}_\perp) + \tau^2 E_\eta({\bf k_\perp}) E_\eta(-{\bf k}_\perp)\right]
\over {\rm Tr}\left[ A_i({\bf k}_\perp) A_i (-{\bf k}_\perp) +
{\tau}^{-2} A_\eta({\bf k}_\perp) A_\eta(-{\bf k}_\perp)\right]}}.$$\,
Remarkably, a mass-gap, which we associate with the
plasmon mass $\omega_{\rm pl} \equiv \omega({\bf k}_\perp =0)$, is generated. 
After initial transient behavior, it satisfies the relation
$$\omega_{\rm pl}(\tau) = \kappa_0 \,g^2\mu \,\sqrt{\frac{1}{g^2\mu\tau}}.$$
The time evolution of this plasmon mass in units of $\omega_{\rm pl} (\tau/g^2\mu)^{1/2}$ 
is shown in Fig.~\ref{fig:plasmon_tau} for $g^2\mu L = 22.6$, with
$\kappa_0 = 0.3\pm 0.01$; it  is 
robust as one approaches the continuum limit. The dependence of
$\omega_{\rm pl}$ on $g^2\mu\, L$ is shown in
Fig.~\ref{fig:plasmon_g2muL}~\footnote{The mass gap was noted in Ref.~\cite{gluodynamics} in context of infrared finite number distributions. Interestingly, the behavior of $\omega_{\rm pl}$ versus $g^2\mu L$ is similar to that of the gluon liberation factor $f_N$. }.

We now assume the growth rate in the expanding system is modified as
$\gamma_{stat} \tau \rightarrow \gamma(\tau) \tau$, and further assume, a) $\gamma(\tau) = m_\infty(\tau)/\sqrt{2}$, and 
b) $m_\infty^2(\tau) = \frac{3}{2}\,\omega_{\rm pl}^2(\tau)$, as in
the static case. Since the plasmon mass is determined independently, these HTL relations predict the Fourier amplitude 
grows as $\exp\left(\Gamma_{th.} \sqrt{g^2\mu \tau}\right)$, with $\Gamma_{th.} = \sqrt{3}\, 
\kappa_0$. This HTL motivated value of the growth rate is compared to the growth rate extracted directly from our fits in 
the table for several values of
$g^2\mu L$ (computed for $N_\perp=32,64$ and $128$, respectively). Remarkably, we find an agreement to within $4-6\%$ accuracy.
\begin{table}[h]
\begin{center}
\begin{tabular}{|c|c|c|c|}
\hline
{\bf $g^2 \mu L$} & $\kappa_0$  & $\Gamma_{\rm th.}=\sqrt{3}\,
\kappa_0$ &
$\Gamma_{\rm fit}$\\
\hline
22.6 & $0.304\pm0.002$  & $0.526\pm0.003$ & $0.502 \pm 0.01$ \\
\hline
67.9 & $0.258\pm0.002$  & $0.447\pm0.003$ & $0.427 \pm 0.01$\\
\hline
90.5 & $0.283\pm0.003$  & $0.49\pm0.004$ & $0.46  \pm 0.04$\\
\hline
\end{tabular}
\end{center}
\end{table}
However, a consistent treatment gives the growth rate in the expanding
case be $\exp( 2 \int_0^\tau  d\tau^\prime \gamma (\tau^\prime) )$. If the HTL relation of $\gamma$ to the plasmon mass is 
unchanged, an additional factor of 2 is obtained in $\Gamma_{th.}$ relative to $\Gamma_{fit}$.
It is not obvious that HTL relations generalize to the 
case of CGC initial conditions. In particular, studies are in progress to investigate
how our measured $\omega_{pl}$ relates to the HTL plasmon frequency. 

Further refinements such as the approach to the continuum limit in the $\eta$ direction require larger lattices. They are 
especially important for studies where the scale $\Delta$, governing
violations of boost invariance, is increased. Much larger values of
$\Delta$ are required to study whether isotropization of particle distributions occurs through classical collective instabilities, a combined instability/modified bottom up scenario~\cite{MSW} or non-perturbative scenarios~\cite{Kovchegov}. 

To summarize, we performed 3+1-D numerical simulations of  SU(2) Yang-Mills equations with CGC initial conditions 
that describe the pre-equilibrium stage of high energy heavy ion collisions. Violations of strict boost invariance cause a non--Abelian Weibel instability. We demonstrated unambiguously that amplitudes of soft modes grow as $\exp(\Gamma
\sqrt{g^2\mu \tau})$ for gluonic matter expanding into the vacuum at
the speed of light. This behavior differs from the $\exp(\sim \tau)$
growth of instabilities in a stationary plasma. The scale governing the growth rate is determined by a plasmon mass generated 
by the non-Abelian dynamics. Extrapolating HTL motivated assumptions to an expanding system,  we compared the growth rate determined from the plasmon mass to the growth rate extracted directly in our simulations. Analytical and numerical work is in progress to test these assumptions. The possible isotropization of distributions due to collective instabilities arising from larger violations of boost invariance is also under investigation.

RV's research is supported by DOE Contract No. DE-AC02-98CH10886. He thanks the Alexander von Humboldt
foundation for support during the early stages of this work. PR was supported by DFG-Forschergruppe EN164/4-4. We thank D.~B\"odeker, J.~Engels, F. Gelis, D. Kharzeev, A. Krasnitz, M.~Laine,
T. Lappi, L. McLerran, Y. Nara and  M.~Strickland for fruitful discussions.

\end{document}